\def\vereq#1#2{\lower3pt\vbox{\baselineskip1.5pt \lineskip1.5pt
\ialign{$\m@th#1\hfill##\hfil$\crcr#2\crcr\sim\crcr}}}

\makeatother
\documentstyle[preprint,12pt,aps]{revtex}

\draft

\title{Comment on 'A new method to calculate the spin-glass order
parameter of the two-dimensional $\pm J$ Ising model'}

\author{ Takayuki {\sc Shirakura}, Fumitaka {\sc Matsubara}$^1$, 
and Michinori {\sc Shiomi}$^1$}

\address{Faculty of Humanities and Social Sciences, Iwate University, 
Morioka 020-8550, Japan \\
$^1$Department of Applied Physics, Tohoku University, Sendai 980-8579,
Japan}

\date{ \today }

\begin{document}

\maketitle

\begin{abstract}
Contrary to the suggestion of
Kitatani and
Sinada (2000 J. Phys. A: Math. Gen. {\bf 33} 3545-3553),
the scaling analysis of the order
parameter distributions suggests the existence of a 
finite-temperature spin-glass phase transition $T_c\ne 0$,
even if the corrections to scaling are taken into account.
\end{abstract}

\pacs{75.50.Lk,75.40.Mg,75.10.Hk}

\sloppy
\input epsf





\newpage

	Kitatani and Sinada\cite{kitatani} suggested that
, in the $\pm J$ Ising model in two dimensions,
if one examines
the scaling analyses considering the corrections to scaling,
every temperature for $0\le T\le 0.3J$ might be able to become the
critical temperature. The suggestion is incompatible with
the recent results of Shirakura et al.\cite{shira1,shira2,shira3},
which suggest the critical temperature $T_c\sim 0.23J$.
However, Kitatani and Sinada performed only the scaling analyses
of the
Binder parameter $g_L$ and the spin-glass susceptibility $\chi_{SG}$. 
In this comment, we perform the scaling
analysis
of the order parameter distribution $P_L(q)$,
using the recent data of the 
authors\cite{siomi}. 

	We can easily show that the following scaling form with
the corrections to scaling
\begin{eqnarray}
	P_L(q) = \frac{L^{\eta /2}}{(1+c/L^\omega )^2}\overline{P}(
	\frac{qL^{\eta /2}}{1+c/L^\omega }) \hspace{1cm}  at
	\hspace{1cm}  T=T_c
\end{eqnarray}
leads to the scaling forms of $g_L$ and $\chi_{SG}$ given by
Kitatani and Sinada\cite{kitatani} with $c=-0.3$ and $\omega =0.5$
(see the eqs.(16) and (17) in Ref. [1]), where $\overline{P}(x)$ 
is some scaling function.
There could be different forms which lead to the same scaling
forms.
For example,
\begin{eqnarray}
	P_L(q) = L^{\eta /2}(1+d/L^\omega )\overline{P}(
	qL^{\eta /2}(1+d/L^\omega )^{1/2}) \hspace{1cm} at 
	\hspace{1cm}  T=T_c
\end{eqnarray}
with $d = 0.6$, also leads to the same scaling forms within
the first order of $1/L^{\omega}$.

	Here we examine the possibility of $T_c=0$.
In figures 1 and 2, we show the scaling plots
corresponding to the eqs. (1) and (2), using the data at the
lowest temperature $T=0.1J$ \cite{comm} in Ref. [5]. 
The difference between 
these two plots is rather small. Therefore we think that
any scaling form of $P_L(q)$ which reproduces the eqs.(16) and (17)
in Ref. [1]
would not change the quality of the plots.
These figures should be compared with the scaling plot 
in figure 10 in Ref. [5], where $T_c=0.23J$ is assumed.
It is evident
that the latter is better than the formers.

	The above result reveals that, even if 
one considers the corrections to scaling, 
the scaling analyses favour the assumption of 
$T_c\sim 0.23J$, not $T_c=0$ in the two-dimensional
$\pm J$ Ising model.

\bigskip
\bigskip
\bigskip







\newpage

\begin{figure}
\caption{The scaling plot of the order parameter distributions 
$P_L(q)$, assuming the form eq. (1) with the corrections to scaling
and using the data at the lowest temperature $T=0.1J$ in Ref. [5]. }
\end{figure}

\begin{figure}
\caption{The scaling plot of the order parameter distributions 
$P_L(q)$, assuming the form eq. (2) with the corrections to scaling
and using the data at the lowest temperature $T=0.1J$ in Ref. [5]. }
\end{figure}


\begin{references}
\bibitem{comm} We think that $P_L(q)$ for $T \rightarrow 0$ may be well 
approximated by that at $T = 0.1J$, because at low temperatures it changes 
a very little as the temperature is decreased(see Fig. 5 in Ref. [5]). 
In particular, 
their peak heights for the smaller lattices($L = 12, 16$) are almost 
unchanged for $T \leq 0.2J$. A slight increase of it is seen for larger 
lattices. If this tendency lasts down to $T = 0$, it will enhance 
the discrepancy of the scaling plot of Figs. 1 and 2.
\end{references}
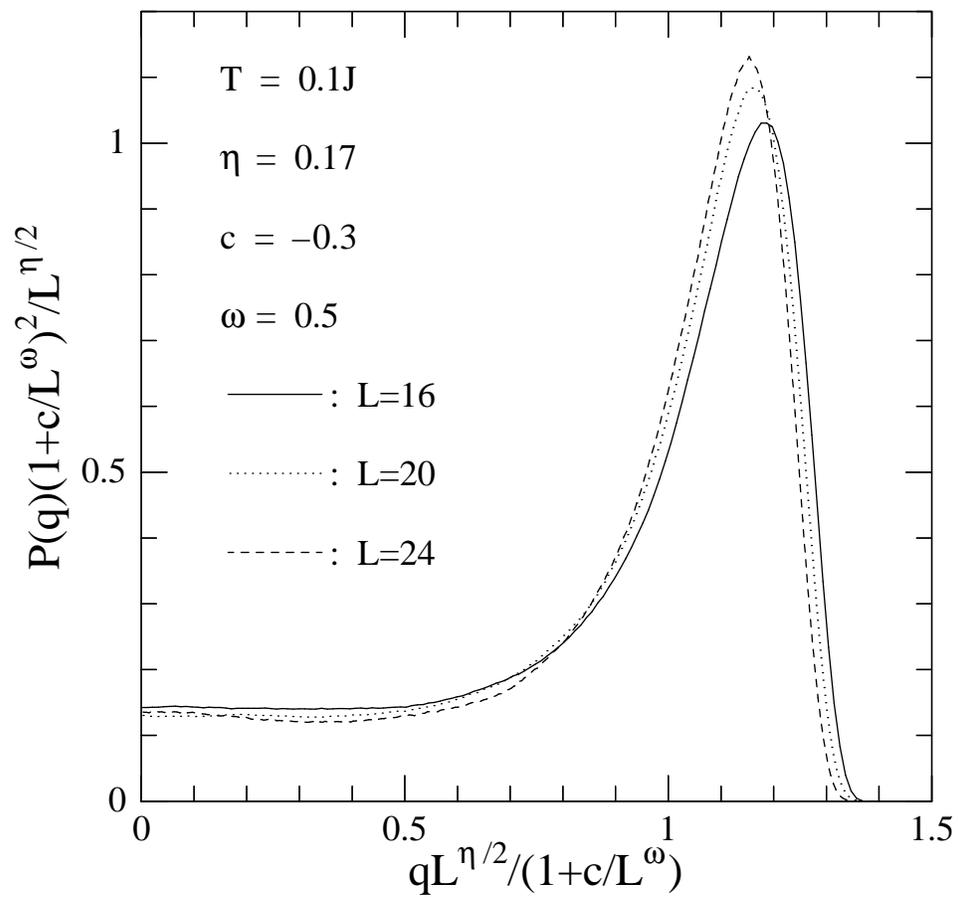
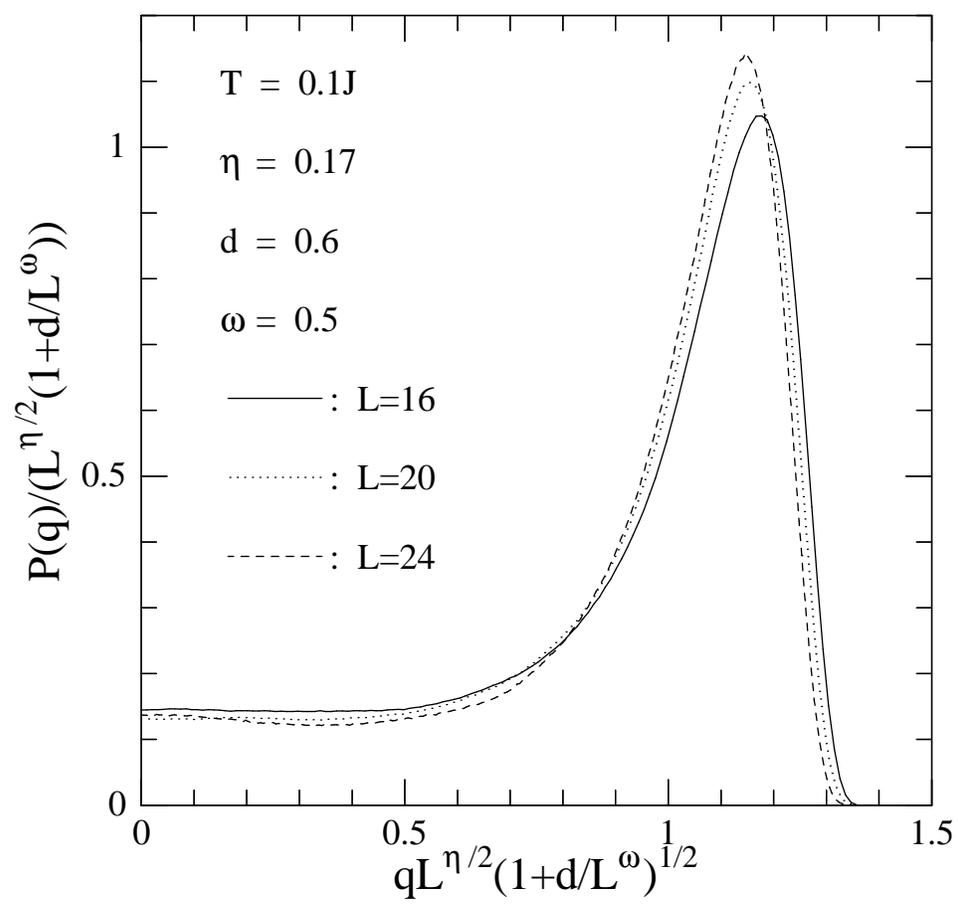
\end{document}